\documentclass[aps,preprint]{revtex4}%
\usepackage{amsfonts}
\usepackage{amsmath}
\usepackage{amssymb}
\usepackage{graphicx}
\usepackage{pdfpages}
\usepackage{xcolor}
\usepackage{caption}
\usepackage{lineno,hyperref}
\usepackage{changes}%
\hypersetup{colorlinks =true, allcolors = blue}
\providecommand{\U}[1]{\protect\rule{.1in}{.1in}}

\begin{document}
\preprint{HEP/123-qed}
\title{Greybody Radiation of scalar and Dirac perturbations of NUT Black Holes}
\author{Ahmad Al-Badawi}
\email{ahmadbadawi@ahu.edu.jo}
\affiliation{Department of Physics, Al-Hussein Bin Talal University, P. O. Box: 20, 71111,
Ma'an, Jordan.}
\author{Sara Kanzi}
\email{sara.kanzi@emu.edu.tr}
\affiliation{}
\author{}
\affiliation{}
\author{\.{I}zzet Sakall{\i}}
\email{izzet.sakalli@emu.edu.tr}
\affiliation{Physics Department, Eastern Mediterranean
University, Famagusta, North Cyprus via Mersin 10, Turkey.}
\author{}
\affiliation{}

\begin{abstract}
We consider the spinorial wave equations, namely the Dirac and the Klein-Gordon equations,
and greybody radiation in the NUT black hole spacetime. To this end, we first study the Dirac
equation in NUT spacetime by using a null tetrad in the Newman-Penrose (NP)
formalism. Next, we separate the Dirac equation into radial and angular sets. The angular set is solved in terms of associated
Legendre functions. With the radial set, we obtain the decoupled radial wave equations and derive the one-dimensional  Schrödinger wave equations together with effective potentials. Then, we discuss the potentials by plotting
them as a function of radial distance in a physically acceptable region. We
also study the Klein-Gordon equation to compute the greybody
factors (GFs) for both bosons and fermions. The influence of the NUT
parameter on the GFs of the NUT spacetime is investigated in detail.

\end{abstract}
\volumeyear{ }
\eid{ }
\date{\today}
\received{}

\maketitle
\tableofcontents
\section{Introduction}

NUT spacetimes are special case of the family of electrovac spacetimes of Petrov type-D given by the Pleba\'{n}ski-Demia\'{n}ski class of solutions \cite{1}. These axially symmetric spacetimes are characterized by the seven
parameters: mass, rotation, acceleration, cosmological constant, magnetic
charge, electric charge, and NUT parameter. We are interested in the solution, which is characterized by two physical parameters: the gravitational mass and the NUT parameter. From now on, we will refer to this spacetime [see metric (\ref{a1})] as the NUT black hole (BH). In fact, the NUT BH is a solution to the vacuum Einstein equation that was found by Newman--Unti--Tamburino (NUT) in 1963 \cite{2}. 

Since the discovery of the NUT solution, its physical meaning is a matter of
debate. The interpretations of the NUT solution and thus the NUT parameter are
still a controversy and it has attracted lots of attention over the years.
Misner approach \cite{3} to the interpretation of the NUT solution aims at
eliminating singular regions at the expense of introducing the
\textquotedblleft periodic\textquotedblright\ time and, consequently, the $%
S^{3}$ topology for the hypersurfaces $r=const$. In \cite{4} the NUT solution was
interpreted as representing the exterior field of a mass located at the
origin together with a semi-infinite massless source of angular momentum.
Whereas in \cite{5} the solution was interpreted as the presence of two
semi-infinite counter-rotating singular regions endowed with negative masses
and infinitely large angular momenta.

Regarding to the NUT parameter, in general it was interpreted as a gravitomagnetic
charge bestowed upon the central mass \cite{6}. However, the exact physical
meaning has not been attained yet. In \cite{7} it was explained that if the NUT
parameter dominates the rotation parameter, then this leaves the spacetime
free of curvature singularities. But if the rotation parameter dominates the NUT
parameter, the solution is Kerr-like, and a ring curvature singularity
arises. Conversely, exact interpretation of the NUT parameter becomes
possible when a static Schwarzschild mass is immersed in a stationary,
source free electromagnetic universe \cite{8}. In this case, the NUT parameter is
interpreted as a twist parameter of the electromagnetic universe. However, in the
absence of this field, it reduces to the twist of the vacuum spacetime. A
different interpretation of the NUT spacetime has been clarified via its
thermodynamics. In \cite{9}, it was shown that if the NUT parameter is interpreted
as possessing simultaneously rotational and electromagnetic features then the thermodynamic quantities satisfy the
Bekenstein-Smarr mass formula.

In article \cite{10}, Misner called the NUT metric a 'counter-example to almost
anything': it means that it has so many peculiar properties. Following this line of thought, the aim of the present
paper is to study the scalar and Dirac perturbations of the NUT BH and analyze their corresponding greybody radiation. 
Studies of this type have always been important in the field of quantum gravity. Furthermore, an
effective way to understand and analyze characteristics of a spacetime is to
study its behavior under different kinds of perturbations including spinor and gauge fields. Also, one can pave the way to study the GFs
of the NUT spacetime in order to investigate the evolution of perturbations
in the exterior region of the BH \cite{11,12}. The
analytical expressions of the solution obtained in this paper could be
useful for the study of the thermodynamical properties of the spinor fields
in NUT\ geometry.

 In the literature, the scalar/Dirac equation and greybody radiation in BH geometry have been extensively studied. For example, wave analysis of the
Schwarzschild BH was performed in \cite{13,14,15,16,17,18}, for the Kerr BH in \cite{19,20,21,22}, for the Kerr
Taub-NUT BH in \cite{23}, for the  Kerr-Newman AdS BH in \cite{24}, and for the Reissner-Nordsrt\"{o}m de
Sitter BH in \cite{25}. Meanwhile, it is worth noting that the GFs measure
the deviation of the thermal spectra from the black body or the so-called Hawking radiation \cite{26,27}. Essentially, GF is a physical quantity that relates to the quantum nature of a BH. A high
value of GF shows a high probability that Hawking radiation can reach
to spatial infinity (or an observer) \cite{28,29}. Today, there are different methods to compute the GF such as the
matching method \cite{30,31}, rigorous bound method \cite{32}, the WKB approximation
\cite{33,34}, and analytical method for the various of spin fields \cite{35,36,37,38,38n,39,40}.
Here, we shall employ  the general method of semi-analytical bounds for the GFs. To this end, we shall examine the characteristic bosonic and fermionic quantum radiation (i.e., the GF) of NUT BH based on the current studies in the literature (see for example \cite{Harmark:2007jy}) and our recent studies \cite{37,38,38n}. Although the methodology followed is parallel to the procedure seen in  \cite{37,38,38n}, due to the unique characteristics of each BH solution, the calculations have their own difficulties and therefore the results obtained will differ significantly. In those recent studies  \cite{37,38,38n}, for example, we have revealed the effects of quintessence, negative cosmological constant, and Lorentz symmetry breaking parameters on the quantum thermal radiation of black holes. However, in this study, we aim to reveal the effect of the NUT parameter on the GF and thus present a theoretical study to the literature that will contribute to the black hole classification studies according to GF measurements that might be obtained in the (distant) future.

The paper is organized as follows: In the next section, we obtain the Dirac
equation in the NUT spacetime and separate the coupled equations into the angular and
radial parts. In Sect. 3, analytical solutions to the angular equations are sought. We also
study the radial wave equations and examine the behaviors of the effective
potential that emerge in the transformed radial equations. In Sect. 4, we
study the Klein--Gordon equation in the NUT BH spacetime. Then, we derive the
associated effective potential of bosons. Then, we compute the GFs of the NUT BH spacetime for both
fermions and bosons in Sect. 5. We draw our conclusions in
Sect. 6.

\section{Dirac equation in NUT BH spacetime}

The NUT metric is a solution to the vacuum Einstein equation that depends on
two parameters: mass $M$ and NUT parameter $l$ \cite{1}. A detailed
discussion on the NUT spacetime can be found in the famous monograph of Griffiths and
Podolsky \cite{41} and in the specific studies given in \cite{42,43,44}. In spherical coordinates $%
(t,r,\theta ,\phi )$ the NUT BH metric reads

\begin{equation}
ds^{2}=f(r)\left[ dt-2l\cos \theta d\phi \right] ^{2}-\frac{1}{f(r)}%
dr^{2}-\left( r^{2}+l^{2}\right) \left( d\theta ^{2}+\sin ^{2}\theta d\phi
^{2}\right),  \label{a1}
\end{equation}

where 
\begin{equation}
f(r)=1-\frac{2\left( Mr+l^{2}\right) }{r^{2}+l^{2}}>0.
\end{equation}

The distinctive feature of the NUT metric is the presence of the NUT
parameter $l$ that causes the spacetime to be asymptotically non-flat. The
horizons are defined by $f(r)=0$ and hence they will occur at $r_{\pm }=M\pm 
\sqrt{M^{2}+l^{2}}$.

Following the NP formalism, we introduce null tetrads  $(l,n,m,\overline{m})$
to satisfy orthogonality relations: $ l.n=-m.\overline{m}=1.$
Thus, we write the basis vectors of null tetrad in terms of elements of the
NUT geometry as follows

\begin{equation*}
l_{\mu }=(1,-\frac{1}{f\left( r\right) },0,-2l\cos \theta ),
\end{equation*}%
\begin{equation*}
n_{\mu }=\frac{1}{2}(f\left( r\right) ,1,0,-2lf(r)\cos \theta ),
\end{equation*}%
\begin{equation*}
m_{\mu }=\sqrt{\frac{r^{2}+l^{2}}{2}}(0,0,-1,-i\sin \theta ),
\end{equation*}%
\begin{equation}
\overline{m}_{\mu }=\sqrt{\frac{r^{2}+l^{2}}{2}}(0,0,-1,i\sin \theta ),
\label{a2}
\end{equation}

\bigskip where a bar over a quantity denotes complex conjugation and the
dual co-tetrad of Eq. (\ref{a2}) are given by%
\begin{equation*}
l^{\mu }=(\frac{1}{f\left( r\right) },1,0,0),
\end{equation*}%
\begin{equation*}
n^{\mu }=\frac{1}{2}(1,-f(r),0,0),
\end{equation*}%
\begin{equation*}
m^{\mu }=\frac{1}{\sqrt{2\left( r^{2}+l^{2}\right) }}(2il\cot \theta ,0,1,%
\frac{i}{\sin \theta }),
\end{equation*}%
\begin{equation}
\overline{m}^{\mu }=\frac{1}{\sqrt{2\left( r^{2}+l^{2}\right) }}(-2il\cot
\theta ,0,1,\frac{-i}{\sin \theta }).  \label{a3}
\end{equation}

Using the above null tetrad one can find the non-zero spin coefficients in
terms of NUT metric:%
\begin{equation*}
\epsilon =\frac{il}{2\left( r^{2}+l^{2}\right) };\gamma =\frac{1}{4}\left(
f^{\prime }(r)+\frac{ilf(r)}{r^{2}+l^{2}}\right) ;\alpha =-\beta =\frac{%
-\cot \theta }{2\sqrt{2\left( r^{2}+l^{2}\right) }}
\end{equation*}

\begin{equation}
\rho =\frac{-r+il}{r^{2}+l^{2}};\mu =\frac{f(r)}{2}\frac{-r+il}{r^{2}+l^{2}}.
\label{a4}
\end{equation}%
Their corresponding directional derivatives become%
\begin{equation*}
D=l^{\mu }\partial _{\mu }=\frac{1}{f(r)}\frac{\partial }{\partial t}+\frac{%
\partial }{\partial r},
\end{equation*}%
\begin{equation*}
\Delta =n^{\mu }\partial _{\mu }=\frac{1}{2}\frac{\partial }{\partial t}-%
\frac{f(r)}{2}\frac{\partial }{\partial r},
\end{equation*}%
\begin{equation*}
\delta =m^{\mu }\partial _{\mu }=\frac{1}{\sqrt{2\left( r^{2}+l^{2}\right) }}%
(2il\cot \theta \frac{\partial }{\partial t}+\frac{\partial }{\partial
\theta }+\frac{i}{\sin \theta }\frac{\partial }{\partial \phi }),
\end{equation*}%
\begin{equation}
\bar{\delta}=\overline{m}^{\mu }\partial _{\mu }=\frac{1}{\sqrt{2\left(
r^{2}+l^{2}\right) }}(-2il\cot \theta \frac{\partial }{\partial t}+\frac{%
\partial }{\partial \theta }-\frac{i}{\sin \theta }\frac{\partial }{\partial
\phi }),  \label{a5}
\end{equation}

The Dirac equations in the curved spacetime are given by \cite{45}%
\begin{equation*}
\left( D+\epsilon -\rho \right) F_{1}+\left( \overline{\delta }+\pi -\alpha
\right) F_{2}=i\mu _{\ast }G_{1},
\end{equation*}%
\begin{equation*}
\left( \Delta +\mu -\gamma \right) F_{2}+\left( \delta +\beta -\tau \right)
F_{1}=i\mu _{\ast }G_{2},
\end{equation*}%
\begin{equation*}
\left( D+\overline{\epsilon }-\overline{\rho }\right) G_{2}-\left( \delta +%
\overline{\pi }-\overline{\alpha }\right) G_{1}=i\mu _{\ast }F_{2},
\end{equation*}%
\begin{equation}
\left( \Delta +\overline{\mu }-\overline{\gamma }\right) G_{1}-\left( 
\overline{\delta }+\overline{\beta }-\overline{\overline{\tau }}\right)
G_{2}=i\mu _{\ast }F_{1},  \label{a6}
\end{equation}

where $\mu _{\ast }$ is the mass of the Dirac particle.

The form of the Dirac equation suggests that we define the spinor fields as
follows \cite{45}, 
\begin{equation}
F_{1}=F_{1}\left( r,\theta \right) e^{i\left( kt+n\phi \right) },  \notag
\end{equation}%
\begin{equation*}
F_{2}=F_{2}\left( r,\theta \right) e^{i\left( kt+n\phi \right) },
\end{equation*}%
\begin{equation*}
G_{1}=G_{1}\left( r,\theta \right) e^{i\left( kt+n\phi \right) },
\end{equation*}%
\begin{equation}
G_{2}=G_{2}\left( r,\theta \right) e^{i\left( kt+n\phi \right) }.  \label{a7}
\end{equation}%
where $k$ is the frequency of the incoming wave and $n$ is the azimuthal
quantum number of the wave.

Substituting the appropriate spin coefficients (\ref{a4}) and the spinors (%
\ref{a7}) into the Dirac equation (\ref{a6}) , we obtain
\begin{equation*}
\sqrt{r^{2}+l^{2}}\left( \emph{D}-\frac{il}{2\left( r^{2}+l^{2}\right) }%
\right) F_{1}+\frac{1}{\sqrt{2}}\mathit{L}F_{2}=i\mu _{0}\sqrt{r^{2}+l^{2}}%
G_{1},
\end{equation*}%
\begin{equation*}
\frac{f}{2}\sqrt{r^{2}+l^{2}}\left( \mathit{D}^{\dag }+\frac{f^{\prime }}{2f}%
-\frac{il}{2\left( r^{2}+l^{2}\right) }\right) F_{2}-\frac{1}{\sqrt{2}}%
\mathit{L}^{\dag }F_{1}=-i\mu _{0}\sqrt{r^{2}+l^{2}}G_{2},
\end{equation*}%
\begin{equation*}
\sqrt{r^{2}+l^{2}}\left( \emph{D}+\frac{il}{2\left( r^{2}+l^{2}\right) }%
\right) G_{2}-\frac{1}{\sqrt{2}}\mathit{L}^{\dag }G_{1}=i\mu _{0}\sqrt{%
r^{2}+l^{2}}F_{2},
\end{equation*}%
\begin{equation}
\frac{f}{2}\sqrt{r^{2}+l^{2}}\left( \mathit{D}^{\dag }+\frac{f^{\prime }}{2f}%
+\frac{il}{2\left( r^{2}+l^{2}\right) }\right) G_{1}+\frac{1}{\sqrt{2}}%
\mathit{L}G_{2}=-i\mu _{0}\sqrt{r^{2}+l^{2}}F_{1},  \label{a8}
\end{equation}%
where, $\mu _{0}=\sqrt{2}\mu _{\ast }$ and the above operators are given by
\begin{equation*}
\emph{D}=\frac{\partial }{\partial r}+\frac{r}{r^{2}+l^{2}}+\frac{ik}{f},
\end{equation*}%
\begin{equation*}
\mathit{D}^{\dag }=\frac{\partial }{\partial r}+\frac{r}{r^{2}+l^{2}}-\frac{%
ik}{f}
\end{equation*}%
\begin{equation*}
\mathit{L}=\frac{\partial }{\partial \theta}+\left( \frac{1}{2}+2lk\right) \cot \theta +%
\frac{n}{\sin \theta },
\end{equation*}%
\begin{equation}
\mathit{L}^{\dag }=\frac{\partial }{\partial \theta}+\left( \frac{1}{2}-2lk\right) \cot
\theta -\frac{n}{\sin \theta }.
\end{equation}%
Further, we define $F_{1}\left( r,\theta \right) =R_{+1/2}\left( r\right)
A_{+1/2}\left( \theta \right) ,F_{2}\left( r,\theta \right) =R_{-1/2}\left(
r\right) A_{-1/2}\left( \theta \right) ,G_{1}\left( r,\theta \right)
=R_{-1/2}\left( r\right) A_{+1/2}\left( \theta \right) $ ,$G_{2}\left(
r,\theta \right) =R_{+1/2}\left( r\right) A_{-1/2}\left( \theta \right) $
hence, the set (\ref{a8}) can be separated into angular and radial parts as 
\begin{equation}
\sqrt{r^{2}+l^{2}}\emph{D}R_{+1/2}=\left( \lambda +i\mu _{0}\sqrt{r^{2}+l^{2}%
}\right) R_{-1/2},  \label{a11b}
\end{equation}%
\begin{equation}
\frac{f}{2}\sqrt{r^{2}+l^{2}}\mathit{D}^{\dag }R_{-1/2}=\left( \lambda -i\mu
_{0}\sqrt{r^{2}+l^{2}}\right) R_{+1/2},  \label{a11}
\end{equation}

\begin{equation}
\frac{1}{\sqrt{2}}\mathit{L}A_{-1/2}=\lambda A_{+1/2},  \label{a10b}
\end{equation}%
\begin{equation}
\frac{1}{\sqrt{2}}\mathit{L}^{\dag }A_{+1/2}=-\lambda A_{-1/2},  \label{a10}
\end{equation}

\bigskip where $\lambda $ is the separation constant.

\section{Solutions of angular and radial equations}

\subsection{Angular equations}

Angular Eqs. (\ref{a10}) and (\ref{a10b}) can be written as

\begin{equation}
\frac{dA_{+1/2}}{d\theta }+\left( \left( \frac{1}{2}+2lk\right) \cot \theta +%
\frac{n}{\sin \theta }\right) A_{+1/2}=-\lambda A_{-1/2},  \label{as2}
\end{equation}%
\begin{equation}
\frac{dA_{-1/2}}{d\theta }+\left( \left( \frac{1}{2}-2lk\right) \cot \theta -%
\frac{n}{\sin \theta }\right) A_{-1/2}=\lambda A_{+1/2}.  \label{as1}
\end{equation}%
We note that the structure of the angular equations admits the similar solutions for both $A_{+1/2}$ and $A_{-1/2}$. Therefore, it is sufficient to focus on one of the angular equations. To this end, we apply the following transformations
\begin{equation}
A_{+1/2}\left( \theta \right) =\cos \left( \frac{\theta }{2}\right) S_{1}(\theta)+\sin
\left( \frac{\theta }{2}\right) S_{2}(\theta),
\end{equation}
\begin{equation}
A_{-1/2}\left( \theta \right) =-\sin \left( \frac{\theta }{2}\right) S_{1}(\theta)+\cos
\left( \frac{\theta }{2}\right) S_{2}(\theta).
\end{equation}
By setting $x=\cos \theta $, one can write Eq. (\ref{as2}) as a decoupled second
order differential equation:

\begin{equation}
\left( 1-x^{2}\right) \frac{d^{2}S_{1}}{dx^{2}}-2x\frac{dS_{1}}{dx}+\left[
\lambda \left( \lambda +1\right) -\frac{\left( n+\frac{1}{2}\right) ^{2}}{%
1-x^{2}}\right] S_{1}=0,  \label{ag1}
\end{equation}

with $\left( n+\frac{1}{2}\right) ^{2}\leq \lambda ^{2}$. The solution to
Eq. (\ref{ag1}) can be expressed in terms of the associated Legendre
functions \cite{46}: 
\begin{equation}
P_{\lambda }^{n}\left( x\right) =\left( 1-x^{2}\right) ^{\tau /2}\frac{d^{n}%
}{dx^{n}}P_{\lambda }\left( x\right) .
\end{equation}

\subsection{Radial equations}

The radial equations (\ref{a11b}-\ref{a11}) can be rearranged as
\begin{equation}
\sqrt{r^{2}+l^{2}}\left( \frac{d}{dr}+\frac{r}{r^{2}+l^{2}}+\frac{ik}{f}%
\right) R_{+1/2}=\left( \lambda -i\mu _{0}\sqrt{r^{2}+l^{2}}\right) R_{-1/2},
\label{a13b}
\end{equation}%
\begin{equation}
\frac{f}{2}\sqrt{r^{2}+l^{2}}\left( \frac{d}{dr}+\frac{r}{r^{2}+l^{2}}+\frac{%
f^{\prime }}{2f}-\frac{ik}{f}\right) R_{-1/2}=\left( \lambda +i\mu _{0}\sqrt{%
r^{2}+l^{2}}\right) R_{+1/2}.  \label{a13}
\end{equation}
To obtain the radial equations in the form of one-dimensional Schr\"{o}%
dinger like wave equations, we perform the following transformations
\begin{equation}
P_{+1/2}=\sqrt{r^{2}+l^{2}}R_{+1/2},\qquad P_{-1/2}=\sqrt{\frac{f}{2}}\sqrt{%
r^{2}+l^{2}}R_{-1/2}.
\end{equation}%
Then, Eqs. (\ref{a13b}) and (\ref{a13}) transform to 
\begin{equation}
\left( \frac{d}{dr}+i\frac{k}{f}\right) P_{+1/2}=\sqrt{\frac{2}{f}}\left( 
\frac{\lambda }{\sqrt{r^{2}+l^{2}}}-i\mu _{0}\right) P_{-1/2},  \label{a15b}
\end{equation}%
\begin{equation}
\left( \frac{d}{dr}-i\frac{k}{f}\right) P_{-1/2}=\sqrt{\frac{2}{f}}\left( 
\frac{\lambda }{\sqrt{r^{2}+l^{2}}}+i\mu _{0}\right) P_{+1/2},  \label{a15}
\end{equation}

and defining the tortoise coordinate $r_{\ast }$ as $\frac{dr_{\ast }}{dr}=%
\frac{1}{f}$, Eqs. (\ref{a15b}) and (\ref{a15}) become 
\begin{equation}
\left( \frac{d}{dr_{\ast }}+ik\right) P_{+1/2}=\sqrt{2f}\left( \frac{\lambda 
}{\sqrt{r^{2}+l^{2}}}-i\mu _{0}\right) P_{-1/2}  \label{b16a}
\end{equation}%
\begin{equation}
\left( \frac{d}{dr_{\ast }}-ik\right) P_{-1/2}=\sqrt{2f}\left( \frac{\lambda 
}{\sqrt{r^{2}+l^{2}}}+i\mu _{0}\right) P_{+1/2}  \label{b16b}
\end{equation}%
In order to write Eqs. (\ref{b16a}) and (\ref{b16b}) in a more compact form,
we combine the solutions as $\psi _{+}=P_{+1/2}+P_{-1/2}$ and $\psi
_{-}=P_{+1/2}-P_{-1/2}$ . After making some computations, we end up with a
pair of one-dimensional Schr\"{o}dinger like equations:
\begin{equation}
\frac{d^{2}\psi _{+}}{dr_{\ast }^{2}}+\left( 4k^{2}-V_{+}\right) \psi
_{+}=0,
\end{equation}%
\begin{equation}
\frac{d^{2}\psi _{-}}{dr_{\ast }^{2}}+\left( 4k^{2}-V_{-}\right) \psi
_{-}=0,
\end{equation}%
with the following effective potentials:
\begin{equation*}
V_{\pm }=\frac{2L^{3/2}\sqrt{\Delta }}{D^{2}}\left[ 2L^{3/2}\sqrt{\Delta }%
\pm 3\mu _{0}^{2}r\Delta \pm L\left( r-M\right) \mp \frac{\Delta L}{D}\left(
2\mu _{0}^{2}r\left( r^{2}+l^{2}\right) -\frac{\lambda \mu _{0}\left(
r-M\right) }{2k}+2rL\right) \right] ,
\end{equation*}%
\begin{equation}
L\left( r-M\right) \mp \frac{\Delta L}{D}\left( 2\mu _{0}^{2}r\left(
r^{2}+l^{2}\right) -\frac{\lambda \mu _{0}\left( r-M\right) }{2k}+2rL\right)
],  \label{p1}
\end{equation}%
where%
\begin{equation}\Delta=r^{2}-2Mr-l^{2},
L=\left( \lambda ^{2}+\mu _{0}^{2}r^{2}\right) , D=L\left(
r^{2}+l^{2}\right) -\frac{\lambda \mu _{0}}{4k}.
\end{equation}%
We see that the potentials (\ref{p1}) depend on the NUT\ parameter 
$l$. To obtain the potentials for massless fermions (neutrino), we simply
set $\mu _{0}=0$ in Eq. (\ref{p1}), which yields 
\begin{equation}
V_{\pm } =\frac{2\lambda \sqrt{\Delta }}{\left( r^{2}+l^{2}\right) ^{2}}%
\left[ 2\lambda \sqrt{\Delta }\pm \left( r-M\right) \mp \frac{2\Delta r}{%
r^{2}+l^{2}}\right].   \label{p2} 
\end{equation}%

To study the asymptotic behavior of the potentials (\ref{p1}), we can expand
it up to order $O\left( \frac{1}{r^{3}}\right) $. Thus, the potentials behave as%
\begin{equation}
V_{\pm }\simeq 4\mu _{0}^{2}+8M\mu _{0}^{2}\frac{1}{r}\pm\left( 6l^{2}\mu
_{0}^{2}-2\lambda ^{2}+\frac{\lambda \mu _{0}}{2k}\right) \frac{1}{r^{2}}%
+O\left( \frac{1}{r^{3}}\right) .  \label{p3a}
\end{equation}

From Eq. (\ref{p3a}), we see that the first term corresponds to the constant
value of the potential at the asymptotic infinity. In the second term, the
mass produces the usual monopole type attractive potential. The NUT
parameter in the third term represents the dipole type potential.

\begin{figure}[h]
\centering
\includegraphics[scale=.8]{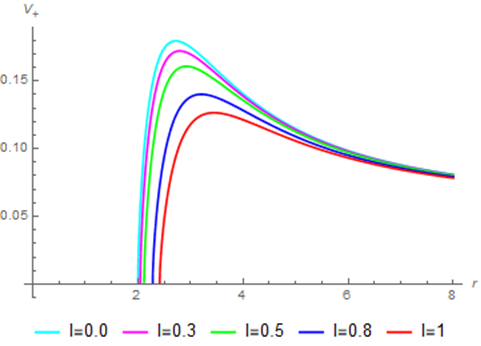}
\caption{Graph of $V_{+}$ (\ref{p1})
for various values of the NUT parameter $l$.  Here, $\lambda = k = 1, \mu = 0.12,$ and $M = 0.5$.}
\label{Figure1}%
\end{figure}
We will now investigate the effect of the NUT parameter on the effective potentials (\ref{p1}) by plotting them as a function of radial distance. Figs. 1 and 2 describe the effective potentials (\ref{p1}) for
massive particles where we obtain potential curves for some specific values
of the NUT parameter. As can be seen from Figs. 1 and 2, for sufficiently small
values of the NUT parameter, the potentials have sharp peaks in the physical
region. When the NUT parameter $l=0$, the peak is seen to be maximum. We
also observe that while the NUT parameter increases, the sharpness of the
peaks decreases, and the peaks tend to disappear after a specific value of
the NUT parameter. This implies that for a large values of the NUT parameter, a
massive Dirac particle moving in the physical region experiences a potential
barrier without peaks and its kinetic energy increases. But for a small
value of the NUT parameter, it feels a sharp potential barrier.

\begin{figure}[h]
\centering
\includegraphics[scale=.8]{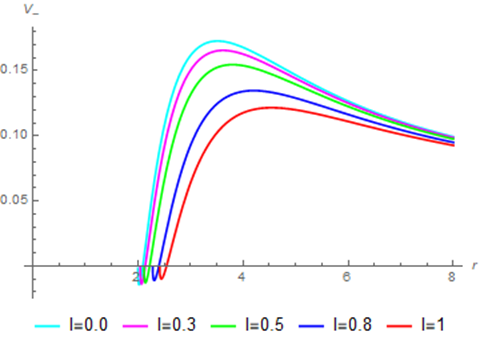} \caption{Graph of $V_{-}$ (\ref{p1})
for various values of NUT parameter $l$. 
 Here, $\lambda = k = 1, \mu = 0.12,$ and $M = 0.5$.}
\label{Figure1}%
\end{figure}

\begin{figure}[h]
\centering
\includegraphics[scale=.8]{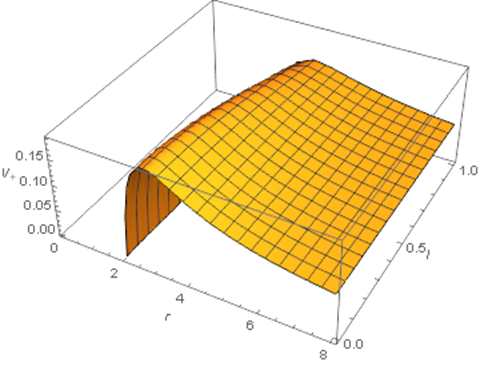} \caption{Three-dimensional plot of $V_{+}$ 
for various values of NUT parameter $l$. Here, $\lambda = k = 1, \mu = 0.12,$ and $M = 0.5$.}
\label{Figure1}%
\end{figure}
Explicitly, the influence of the NUT parameter is shown in the
three-dimensional plots of the potentials with respect to the NUT parameter
and the radial distance. As can be seen from Fig. (3), we observe a
peak for small values of the NUT parameter and as the value of the NUT
parameter increases, the potentials decrease. Let us note that, for massless
Dirac particle $\left( \mu _{0}=0\right)$, the potentials have similar
behaviors as for the massive case.

\section{Greybody Radiation from NUT BH}

\subsection{Klein-Gordon equation}

The Klein-Gordon equation for the massless scalar field in curved coordinates is given by \cite{isepjc} 
\begin{equation}
\square \Psi =\frac{1}{\sqrt{-g}}\partial _{\mu }\sqrt{-g}%
g^{\mu \nu }\partial _{\nu } \Psi =0,  \label{is1}
\end{equation}

where $g$ is the determinant of the spacetime metric (\ref{a1}): $%
\sqrt{-g}=\left( r^{2}+l^{2}\right) \sin \theta $. To separate the variables
in equation (\ref{is1}), we assume the solutions to the wave equation in the
form 
\begin{equation}
\Psi =R\left( r\right) \Theta \left( \theta \right) \exp \left( -i\omega
t\right) \exp \left( im\phi \right) ,
\end{equation}

where $\omega $ denotes the frequency of the wave and $m$ is the azimuthal
quantum number. Therefore, the Klein-Gordon equation (\ref{is1}) becomes%
\begin{equation*}
\left( \frac{\left( r^{2}+l^{2}\right) ^{2}}{\Delta }-\frac{4l^{2}\cos
^{2}\theta }{\sin ^{2}\theta }\right) \omega ^{2}+\frac{1}{R\left( r\right) }%
\frac{d}{dr}\left( \Delta \frac{dR\left( r\right) }{dr}\right) +
\end{equation*}%
\begin{equation}
\frac{1}{\Theta \left( \theta \right) }\frac{1}{\sin \theta }\frac{d}{%
d\theta }\left( \sin \theta \frac{\partial \Theta \left( \theta \right) }{%
\partial \theta }\right) -\frac{m^{2}}{\sin ^{2}\theta }-\frac{4l\cos \theta 
}{\sin ^{2}\theta }\omega m=0.  \label{kg1}
\end{equation}
The angular part satisfies%
\begin{equation}
\frac{1}{\sin \theta }\frac{d}{d\theta }\left( \sin \theta \frac{\partial
\Theta \left( \theta \right) }{\partial \theta }\right) -\left( \frac{%
4l^{2}\cos ^{2}\theta +4l\omega m\cos \theta +m^{2}}{\sin ^{2}\theta }%
\right) \Theta \left( \theta \right) =-\lambda \Theta \left( \theta \right),
\end{equation}%
where it has solutions in terms of the oblate spherical harmonic functions
having eigenvalues $\lambda $ \cite{46}. Therefore, the Klein--Gordon equation (%
\ref{kg1}) is left with the radial part

\begin{equation}
\frac{d}{dr}\left( \Delta \frac{d}{dr}R\left( r\right) \right) +\left(
\left( r^{2}+l^{2}\right) ^{2}\omega ^{2}-\Delta \lambda \right) R\left(
r\right) =0.  \label{is5}
\end{equation}

By changing the variable in a new form as $R\left( r\right) =\frac{U\left(
r\right) }{\sqrt{r^{2}+l^{2}}},$ the radial wave equation (\ref{is5})
recasts into a one-dimensional Schr\"{o}dinger like equation as follows%
\begin{equation}
\frac{d^{2}U}{dr_{\ast }^{2}}+\left( \omega ^{2}-V_{eff}\right) U=0,
\end{equation}%
in which $r_{\ast }$ is the tortoise coordinate: $\frac{dr_{\ast }}{dr}=%
\frac{r^{2}+l^{2}}{\Delta },$ and $V_{eff}$ is the effective potential given
by%
\begin{equation}
V_{eff}=\frac{\Delta }{\left( r^{2}+l^{2}\right) ^{3}}\left( \frac{%
-3r^{2}\Delta }{r^{2}+l^{2}}-\lambda \left( r^{2}+l^{2}\right)
-3r^{2}-4Mr-l^{2}\right) .  \label{v1}
\end{equation}

\begin{figure}[h]
\centering
\includegraphics[scale=.8]{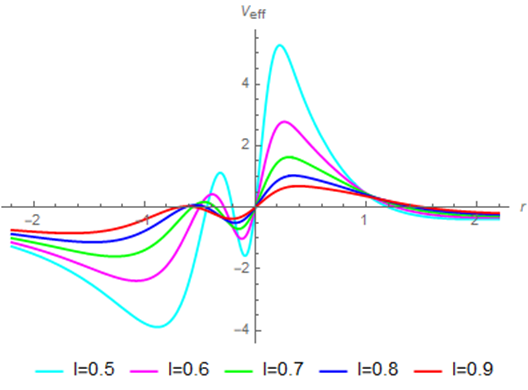} \caption{Plots of $V_{eff}$  (\ref{v1}) versus $r$ for the spin-$0$ particles for various values of NUT parameter $l$. The physical parameters are chosen as; $\lambda = k = 1, \mu = 0.12,$ and $M = 0.5$. 
}
\label{Figure1}%
\end{figure}

It is seen from (\ref{v1}) that the effective potential
depends on the NUT\ parameter $l$. Fig. 4 indicates the behavior of
potential (\ref{v1}) under the effect of NUT parameter for massless
particles. As can be seen from Fig. (4),  when the NUT parameter increases, peaks of the potential goes down.

\subsection{GFs of bosons}

It is known that nothing can escape a BH when approaching it. But when a
quantum effect is considered, the BH can radiate. This radiation is known as
the Hawking radiation. We now assume that Hawking radiation is a
massless scalar field (bosons) that satisfies the Klein--Gordon equation. To evaluate the GF of the NUT metric for bosons, we use \cite{47}
\begin{equation}
\sigma _{l}\left( \omega \right) \geq \sec h^{2}\left( \int_{-\infty
}^{+\infty }\wp dr_{\ast }\right) ,  \label{is8}
\end{equation}

in which $r_{\ast }$ is the tortoise coordinate and 
\begin{equation}
\wp =\frac{1}{2h}\sqrt{\left( \frac{dh\left( r_{\ast }\right) }{dr_{\ast }}%
\right) ^{2}+(\omega ^{2}-V_{eff}-h^{2}\left( r_{\ast }\right) )^{2}},
\label{is9}
\end{equation}

where $h(r\ast )$ is a positive function satisfying $h\left( -\infty \right)
=h\left( -\infty \right) =\omega $: see Refs. \cite{47,48} for more details. Without loss of generality, we select $%
h=\omega $. Therefore,%
\begin{equation}
\sigma_{l} \left( \omega \right) \geq \sec h^{2}\left( \int_{r_{h}}^{+\infty }%
\frac{V_{eff}}{2\omega }dr_{\ast }\right) .  \label{gb1}
\end{equation}%
We use the potential derived in (\ref{v1}) to obtain the GFs for bosons,
namely

\begin{equation}
\sigma^{s}_{l} \left( \omega \right) \geq \sec h^{2}\frac{1}{2\omega }%
\int_{r_{h}}^{+\infty }\frac{1}{\left( r^{2}+l^{2}\right)}\left( \frac{%
-3r^{2}\Delta }{\left( r^{2}+l^{2}\right) ^{2}}+\lambda 
+\frac{\Delta^{\prime }{r}+\Delta}{r^{2}+l^{2}}\right) dr.
\end{equation}%
After integration, we obtain 
\begin{equation}
\sigma^{s}_{l} \left( \omega \right) \geq \sec h^{2}\frac{1}{2\omega }%
{\left(\frac{-\lambda}{r}-\frac{M}{r^{2}}+\frac{l^{2}\left(\lambda-5\right)}{3r^{3}}+\frac{5Ml^{2}}{2r^{4}}+\frac{l^{4}\left(16-\lambda\right)}{5r^{5}}+\frac{l^{6}\left(\lambda-3\right)}{7r^{7}}\right)}. \label{SK2}
\end{equation}
The behaviors of the obtained bosonic GFs of the NUT BH are depicted in Fig. 5
in which the plots of $\sigma \left( \omega \right) $ for various values of NUT\
parameter. As can be seen from those graphs, the NUT has an effect on the bosonic GFs. Remarkably,  $\sigma \left( \omega \right) $ clearly decreases with the increasing NUT
parameter. In other words, the NUT plays a kind of booster role for the
GFs of the spin-0 particles.

\bigskip 

\begin{figure}[h]
\centering
\includegraphics[scale=.5]{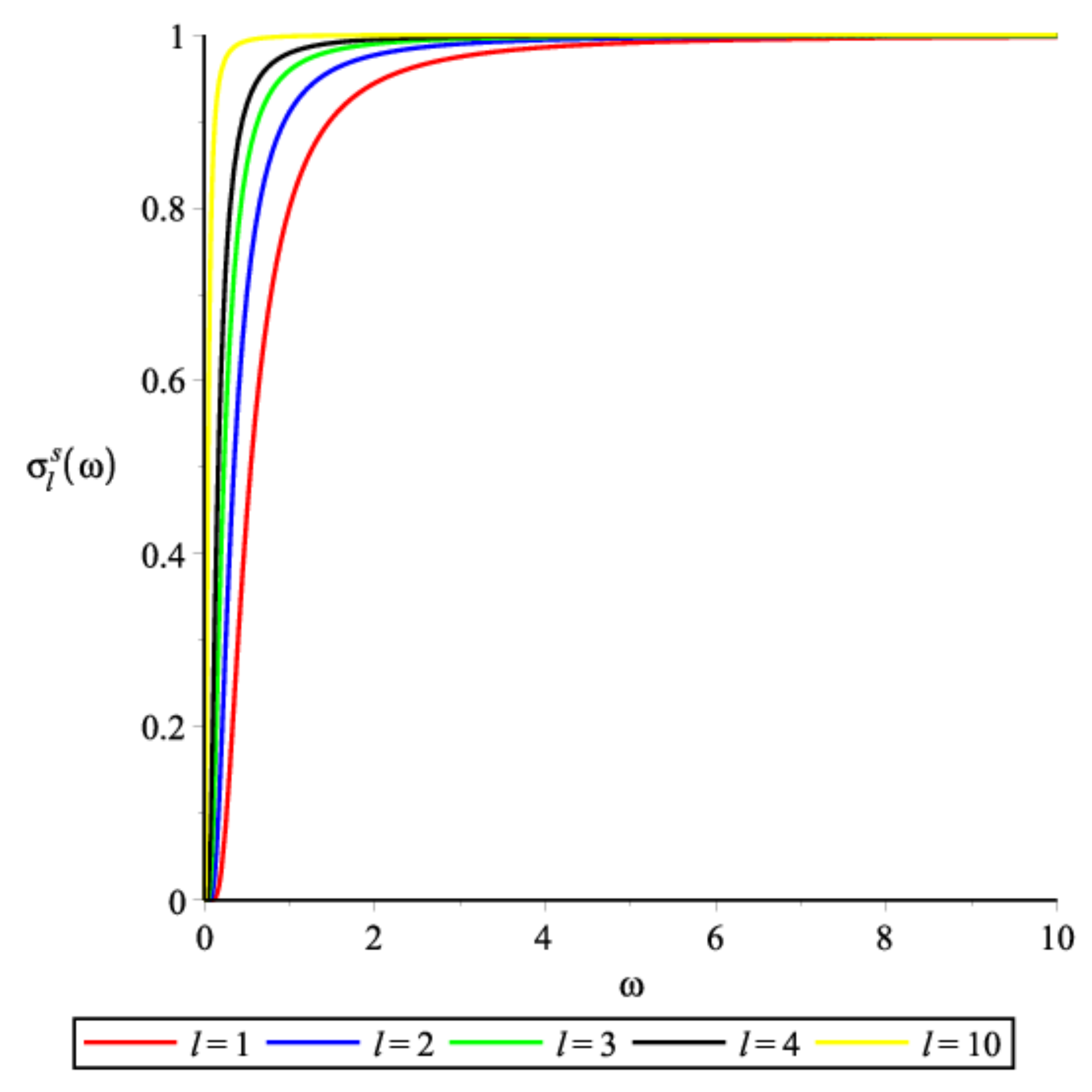}\caption{Plots of $\sigma^{s}_{l}\left(\omega \right)$  (\ref{SK2}) versus $\omega$ for the zero spin particles for various values of NUT parameter $l$. The physical parameters are chosen as; $\lambda =2$
and $M = 1$.}
\label{FigureK2}%
\end{figure}

\subsection{GFs of fermions}

Here, we shall derive the fermionic GF of the neutrinos emitted from the
NUT\ BH. To this end, we consider the potentials (\ref{p2}). Following the
procedure described above [see Eqs. (\ref{is8})-(\ref{gb1})], we obtain,

\begin{equation}
\sigma_{\pm} \left( \omega \right) \geq \sec h^{2}\frac{\lambda}{\omega }%
\left(-\frac{2\lambda\mp1}{r}\mp\left(\frac{M}{r^{2}}+\frac{M^{2}-l^{2}\mp4l^{2}\lambda}{6r^{3}}\right)\pm\frac{3M^{3}-Ml^{2}}{8r^{4}}\right).\label{Kanzi1}
\end{equation}

The fermionic GFs of the NUT BH are depicted in Figs. 6 and 7 in which the plots of $\sigma_{\pm} \left( \omega \right) $ for various values of NUT\
parameter are obtained. Similar to the bosonic GFs, the NUT has an intensifier effect on the
GFs of the spin-$\frac{1}{2}$ particles.

\begin{figure}[h]
\centering
\includegraphics[scale=.5]{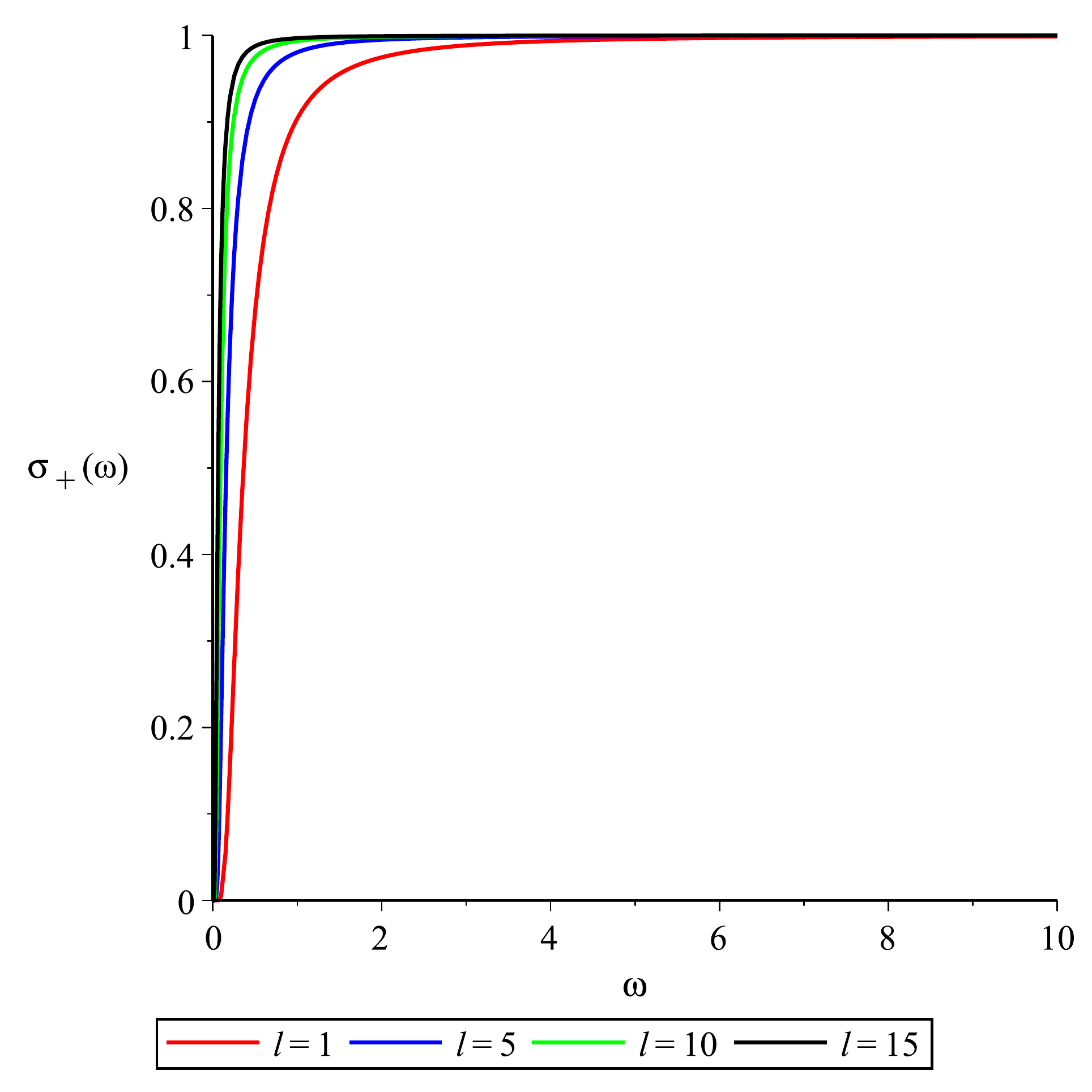}\caption{Plots of $\sigma_{+}\left(\omega \right)$  (\ref{SK2}) versus $\omega$ for the Non-zero spin particles for various values of NUT parameter $l$. The physical parameters are chosen as; $\lambda =-0.5$
and $M = 1$.}
\label{FigureK1}%
\end{figure}

\begin{figure}[h]
\centering
\includegraphics[scale=.5]{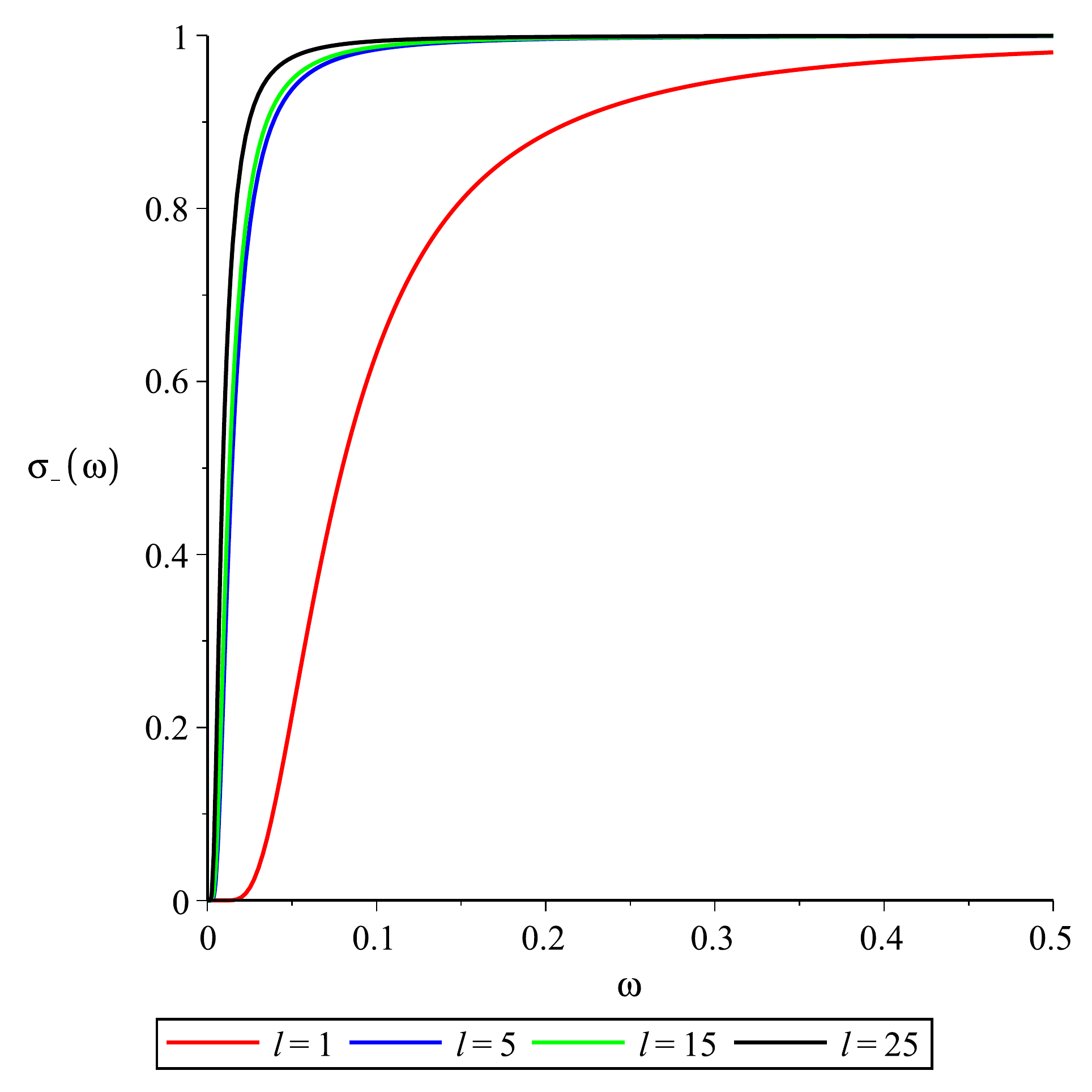}\caption{Plots of $\sigma_{-}\left(\omega \right)$  (\ref{Kanzi1}) versus $\omega$ for the Non-zero spin particles for various values of NUT parameter $l$. The physical parameters are chosen as; $\lambda =-0.5$
and $M = 1$.}
\label{FigureK2}%
\end{figure}

\section{Conclusion}

We have studied the perturbation and greybody radiation{\LARGE \ }in the NUT
spacetime. The NUT metric describes the vacuum spacetime around a source
that is characterized by two parameters, the mass of the central object and
the NUT parameter. The perturbations studied here are for the bosons and fermions,
namely the Klein-Gordon and Dirac equations are considered. For the Dirac equation, we
explicitly work out the separability of the equations into radial and
angular parts by using a null tetrad in the NP formalism. It is shown that
solutions to the angular equations could be given in terms of the associated
Legendre functions. We also discussed the radial equations and obtain a wave
equation with an effective potential. The radial part involves both mass and NUT parameters. It is shown in the asymptotic expansion of the
potentials (\ref{p3a}) that the mass produces the usual $1/r$ attractive
potential while the NUT parameter manifests itself in the $1/r^{2}$
asymptotically. Thus, the effect of gravity is as expected stronger than the
effect of the NUT parameter in the NUT geometry. To understand the physical
interpretations of the potentials (\ref{p1}), we make two- and
three-dimensional plots of the potentials for different values of the NUT
parameter. It is seen from the plots that the potential barriers are higher for
small values of the NUT parameter. If the NUT parameter decreases, the peak
of the potential barriers distinctly increases. Potentials become limited
regardless of the value of the NUT parameter and tend to a constant value
for large values of distances.

For the GFs of bosons, we have considered the massless scalar wave equation.
To this end, we have reduced the radial part of the Klein-Gordon equation to the
one-dimensional Schr\"{o}dinger like wave equation. The behaviors of the
effective potential (\ref{v1}) under the effect of the NUT parameter for the
scalar waves are depicted. It is seen from Fig. 4 that as the NUT parameter
increases then the peak of the barrier goes down. In the sequel, we have computed the
GFs, one of the fundamental information that can be obtained
from the BHs, for both bosons and fermions. It has been observed from Figs. 5, 6 and 7 that increment of the NUT value significantly increases the GF radiation of both bosons and fermions.

In the cosmology, the NUT parameter measures the
amount of anisotropy that a metric has at large times [50]. Therefore, the GF analyses of the rotating NUT BHs, like Kerr-NUT-de Sitter BH [50], in the AdS background shall be a significant future extension of the present work. This may also be important to understand the AdS/CFT conjecture with
the NUT similar to the quasinormal modes [51] analysis of  planar Taub-NUT BHs [52]. This will be
investigated in a near future work.

\end{document}